\documentclass[twocolumn,amssymb,aps,prd]{revtex4}
\usepackage{epsfig}

\begin{document}

\hspace{8cm}\hbox to\hsize{\hfill\tt  PNUTP-04/A01}
 \vspace{0.9cm}

\title{Brane World Confronts Holography
\footnote{Talk delivered by DKH for the International
conference on Flavour physics (ICFP-II) in KIAS, Seoul, Korea,
Oct. 6-11, 2003.}
}%

\author{Deog Ki Hong}%
\email{dkhong@pusan.ac.kr} \affiliation{Department of
Physics, Pusan National University,  Pusan 609-735, Korea}
\author{Stephen D.~H. Hsu}%
\email{hsu@duende.uoregon.edu} \affiliation{
Department of Physics,
University of Oregon, Eugene OR 97403-5203, U.S.A.}

\begin{abstract}
Holography principle imposes a stringent constraint on the scale of
quantum gravity $M_*$ in brane-world scenarios, where all matter is confined
on the brane.
The thermodynamic entropy
of astrophysical black holes and sub-horizon volumes during big
bang nucleosynthesis exceed the relevant bounds unless $M_* >
10^{(4-6)}$ TeV, so a hierarchy relative to the weak scale is
unavoidable. We discuss the implications for extra dimensions as
well as holography.
\end{abstract}

\maketitle 

\newpage

\section{Introduction}
The idea that our universe might have extra dimensions beyond
four space-time dimensions dates long back to early last
century.
In 1921 Kaluza~\cite{Kaluza:tu} and later Klein~\cite{Klein:tv}
introduced extra dimensions to unify
all forces in nature. The metric for higher dimensions was
postulated to be
\begin{eqnarray}
\hat g_{\hat\mu\hat\nu} &=&
\left(
\begin{array}{cc}
g_{\mu\nu}-\varphi A_{\mu}A_{\nu} \quad \varphi A_{\mu} \\
\varphi A_{\nu}  \quad \quad\quad\varphi
\end{array}\right)\,,
\end{eqnarray}
where $g_{\mu\nu}$ is the ordinary 4d metric and
$A_{\mu}$ is interpreted as the photon field.
After the dimensional reduction to $M^4\times S^1$, the general
coordinate transformation induces the $U(1)$ gauge transformation in
$M^4$.

String theory proposed as the theory of everything
is consistent only in 10 dimensions, where 6
extra dimensions are compactified to be a Calabi-Yau
manifold~\cite{Gross:1984dd,Candelas:en}.

Soon after Horava-Witten~\cite{Horava:1995qa}
pointed out that our world may be confined
on a brane embedded in eleven dimensional spacetime, a low-scale gravity
or a brane-world scenario~\cite{Arkani-Hamed:1998rs,Randall:1999ee}
is proposed as a solution to the gauge hierarchy problem. Letting gravity
propagate in extra dimensions while all standard model particles are
confined on a brane, the scale of gravity can be made arbitrary.
\begin{figure}
\begin{center}
\leavevmode \epsfig{file=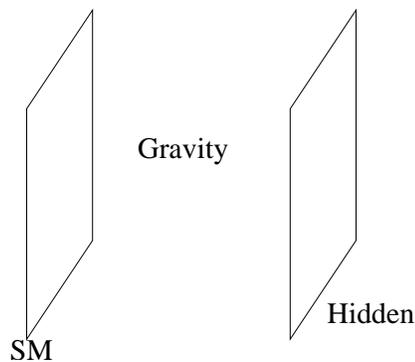,width=0.3\textwidth
} \caption{ Only gravity propagates in bulk. }
\label{fig:rs}
\end{center}
\end{figure}
In the brane world the Einstein action has the form
\begin{equation}
S = \int
d^4x~ M_*^2 {\cal R} ~\left( d^{D-4}x ~ M_*^{D-4} \right)~ \sqrt{-g} ~~.
\end{equation}
The relation between the fundamental Planck scale $M_*$ and the
apparent one $M_P$, the four-dimensional Planck scale, is given by
\begin{equation}
V_w M_*^{D-2} = M_P^2\,,\quad V_w\equiv
\int d^{D-4}x ~\sqrt{-g_{(D-4)}}\,.
\end{equation}
By adjusting $V_w$, the volume of the extra dimensions,
we may take  $M_* = O(1)~ {\rm TeV}$ to solve the hierarchy problem.
Since this low-scale gravity will be different from the usual
gravity at short scales, Newton's law deviates at sub millimeter.
The model is soon to be tested. However, in this talk we will argue that
the holography bound requires the fundamental Planck scale can not be too
small~\cite{Hong:2003xd}.
To reproduce the successful nucleosynthesis in Big Bang cosmology
and to account for the supernova explosion, $M_*>10^{4-6}~{\rm TeV}$ and
the brane-world solution thus has a little hierarchy problem, even if it
is operating.

\section{What is Holography Bound}
Bekenstein~\cite{Bekenstein:ur,Bekenstein:ax} conjectured that
for a system of energy
$M$ in a radius $R$, its entropy is bounded from above
\begin{equation}
S<{2\pi M R\over \hbar}.
\end{equation}
For weak gravity, the size of the system is much larger than its
Schwarzschild radius, $R_s~(=2GM)<R$. Therefore we get
\begin{equation}
S<{2\pi M R\over \hbar}<{A\over 4G\hbar},~~\quad A=4\pi R^2.
\end{equation}
The entropy of a system is less than one quarter of its area
in the unit of Planck area, $l_p^2=G\hbar/c^3$.

The Bekenstein bound for entropy lead 't~Hooft~\cite{'tHooft:gx}
and Susskind~\cite{Susskind:1994vu} to
formulate the holography principle, which states that
the entropy in a spatial volume $V$ enclosed by a surface area $A$
cannot exceed $A/4$ in Planck units. Consider a system in a box.
In quantum field theory, a state in a box can have arbitrarily large energy.
However, if its compton length is smaller than its Schwarzschild radius,
an observer outside the box can not access such a state. Therefore, the energy
of states in the box is limited to outside observers.
Such states do not contribute to entropy of the system measured by outside
observers. For outside observers, the number of accessible states of the system
is much less than that of the states allowed by a local quantum field theory.

In $D$ dimensions, the Schwarzschild radius $R_s$ of a system with energy $E$
is determined roughly by the condition that the gravitational potential
energy is of order one at $R=R_s$. If $R_s$ is smaller than the radius of the
extra dimensions,
\begin{equation}
\Phi ~\sim~ {E \over M_*^{D-2}
R^{D-3}} ~\longrightarrow~
R_s \sim ( M_*^{2-D} E )^{1/(D-3)}.
\end{equation}
Therefore, the energy of a system of size $R$ must have a upper bound not
to collapse into a black hole. If $E_{\rm max}$ is the maximum energy of the
system, then $E_{\rm max} <a^{-D}R^{D-1}$, where $a^{-1}$ is the ultraviolet
(UV) cutoff. Not to collapse into a black hole, the size of the system has to
be bigger than its Schwarzschild radius
\begin{eqnarray}
( M_*^{2-D}\! E_{\rm max} )^{\!1/(D-3)}\!<\!
( M_*^{2-D}\! a^{-D}R^{D-1} )^{\!1/(D-3)}\!<\!R\,.
\end{eqnarray}
We find that the UV cutoff is related to the infrared cutoff of the system,
\begin{equation}
a>M_*^{-1}\,\left(R\,M_*\right)^{2/D}.
\end{equation}

The entropy of the system~\cite{Cohen:1998zx} is given as ($k_B=1$),
neglecting all quantum numbers except positions,
\begin{equation}
S=\ln 2^{(R/a)^{D-1}}<\left(RM_*\right)^{D-3+2/D}\ln2\,.
\end{equation}
For $D=4$, this bound gives $S<C\,A^{3/4}$, which is smaller than
the area $A$.

\section{Covariant Entropy Bound}
\begin{figure}[h]
\epsfxsize=1.2in
{\centerline{\epsffile{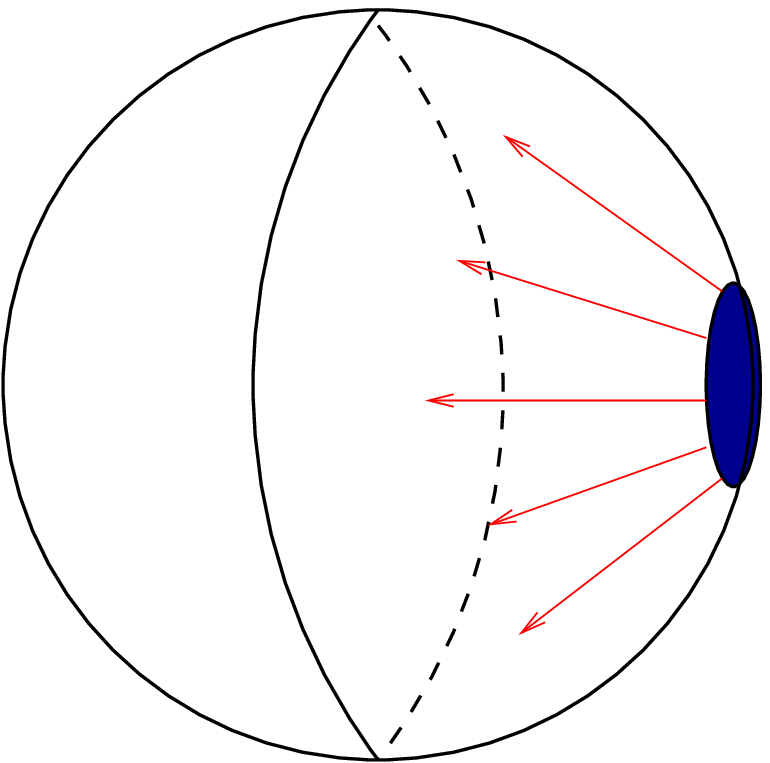}\hskip 1in {\epsfxsize=0.8in \epsffile{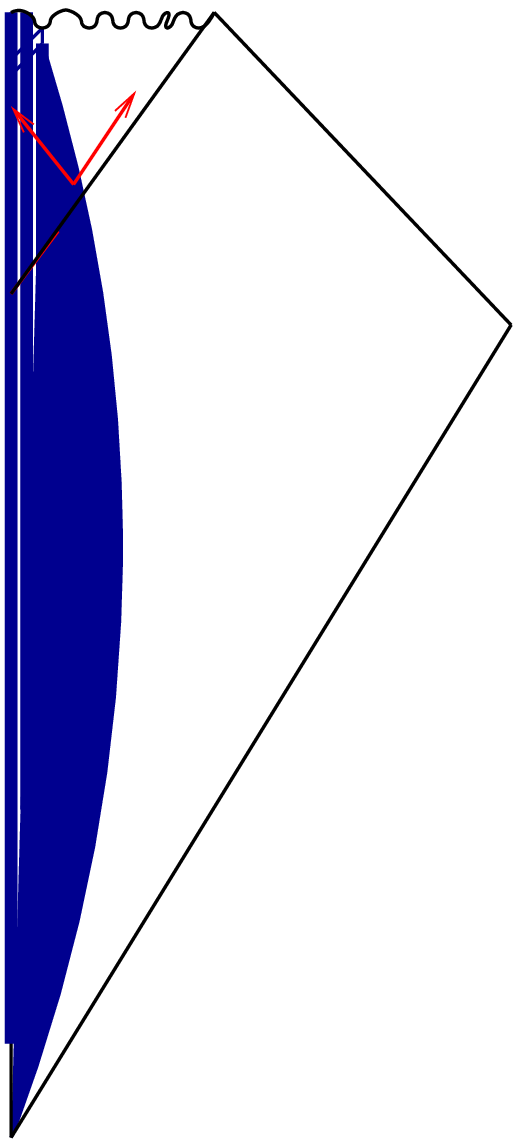}}}}
\caption{Closed universe and a collapsing star}
 \label{cos}
\end{figure}

The Bekenstein bound for entropy applies to a static system only. For instance
one can easily see that a closed universe or a collapsing star, where
the system evolves in time, violates
the Bekenstein bound (See Fig.~{\ref{cos}}).
For a closed universe, near the big crunch, the area of the closed universe
can be made arbitrarily small while its entropy never decreases.
Similarly in the evolution of a collapsing star the spatial area of the
collapsing star becomes arbitrarily small, violating the Bekenstein bound.

\begin{figure}[h]
\epsfxsize=0.8in
{\centerline{\epsffile{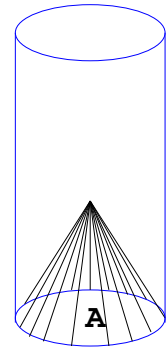}}}
\caption{Entropy on a null hypersurface, $L$.}
 \label{cov}
\end{figure}

Since the spacelike volume does not have an intrinsic meaning in general
relativity, one may introduce an intrinsic entropy bound~\cite{Fischler:1998st}.
In 1999 Bousso~\cite{Bousso:1999xy,Bousso:2002ju}
introduced a covariant entropy bound, which states that the entropy on
any light-sheet of a surface $B$ will not exceed the area of $B$:
\begin{equation}
S_L\le {A(B)\over4}.
\end{equation}
The null geodesics extended from the surface will merge at a focal point
in the future direction~\cite{foot1}
(See Fig.~\ref{cov}), defining a null hypersurface
$L$.
As time evolves, all matter inside $B$ of spatial volume $V$ will pass
through the null hypersurface, $L$.
By the second law of thermodynamics, the entropy in the null
hypersurface is not less than the entropy inside $B$.
Therefore, the covariant entropy bound gives the Bekenstein bound,
\begin{equation}
S_V<S_L<{A\over4}\,.
\end{equation}
Furthermore it does remain valid in the case of  dynamical systems
like collapsing universe or collapsing stars.

\section{Black Hole Thermodynamics}

Original Bekenstein conjecture on entropy bound was motivated by
black hole thermodynamics, where the entropy of black hole is
found to be proportional to the area of the black hole horizon.

Here, we review the derivation of the black hole entropy
by 't~Hooft~\cite{'tHooft:1984re},
which is based on two distinct properties of black holes.
The first property is that black holes radiate as black bodies with a
certain temperature, called Hawking temperature,
\begin{equation}
T_{\rm H}={1\over 8\pi M},
\end{equation}
where $M$ is the mass of the black hole. The second property is that
black holes have an event horizon. If one drops
an object with energy $\Delta E$ into a black hole
with mass $E$. ($\Delta E\ll1\ll E$ in Planck units.)
Then, the absorption cross section is then
\begin{equation}
\sigma=\pi R^2,~\quad~R\simeq 2E.
\end{equation}
From the Hawking's result, the emission probability is
\begin{equation}
W\simeq \pi R^2\rho_{\Delta E}\,e^{-\beta_{\rm H}\Delta E},
\end{equation}
where $\rho_{\Delta E}$ is the density of states for a particle with
energy $\Delta E$.
Now, if we suppose the processes are described by a Hamiltonian acting in Hilbert
space. Then,
\begin{eqnarray}
\sigma&=&\left|\left<E+\Delta E\left|T\right|E,\Delta E\right>\right|^2
\rho(E+\Delta E)\nonumber\\
W&=&\left|\left<E,\Delta E\left|T\right|E+\Delta E\right>\right|^2
\rho(E)\rho_{\Delta E}.
\end{eqnarray}
By $PCT$ invariance, the matrix elements have to be same and we get
\begin{equation}
{\sigma\over W}={e^{\Delta E/T_{\rm H}}\over \rho_{\Delta E}}
={\rho(E+\Delta E)\over \rho(E)\rho_{\Delta E}}
\end{equation}
Therefore, we find the density of states of the black hole
$\rho(E)=\exp(4\pi E^2)$ and the black hole entropy becomes
\begin{equation}
S=\ln \rho(E)=4\pi E^2+S_0\quad {\rm or}\quad S={A\over 4}+S_0.
\end{equation}
where $S_0$ is the subleading term.

In the next section, we will apply the holography bound and the black hole
entropy bound to the brane world scenarios,
both ADD~\cite{Arkani-Hamed:1998rs} and RS models~\cite{Randall:1999ee}.

\section{Holography Bounds on Brane World Scenario}

Consider an ADD/RS world in which the standard model degrees of
freedom are confined to a 3-brane while the gravitational degrees
of freedom propagate in D dimensions. The large effective volume
$V_w$ of the bulk allows the apparent Planck scale $M_P$ to be
much larger than the true dynamical scale of gravity $M_* \sim
{\rm ~TeV} $. 

\bigskip
\noindent (1) {\it The holographic bound is violated during the
big bang.}

\begin{figure}[h]
\centerline{{\epsfxsize=0.7in\epsffile{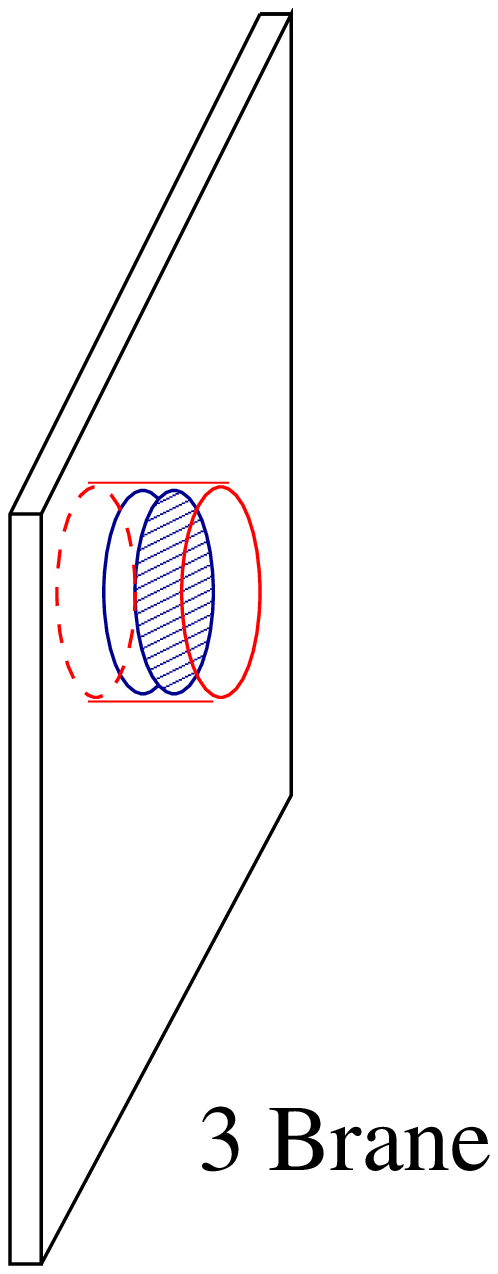}}
\hskip 0.2in {\epsfxsize=2in \epsffile{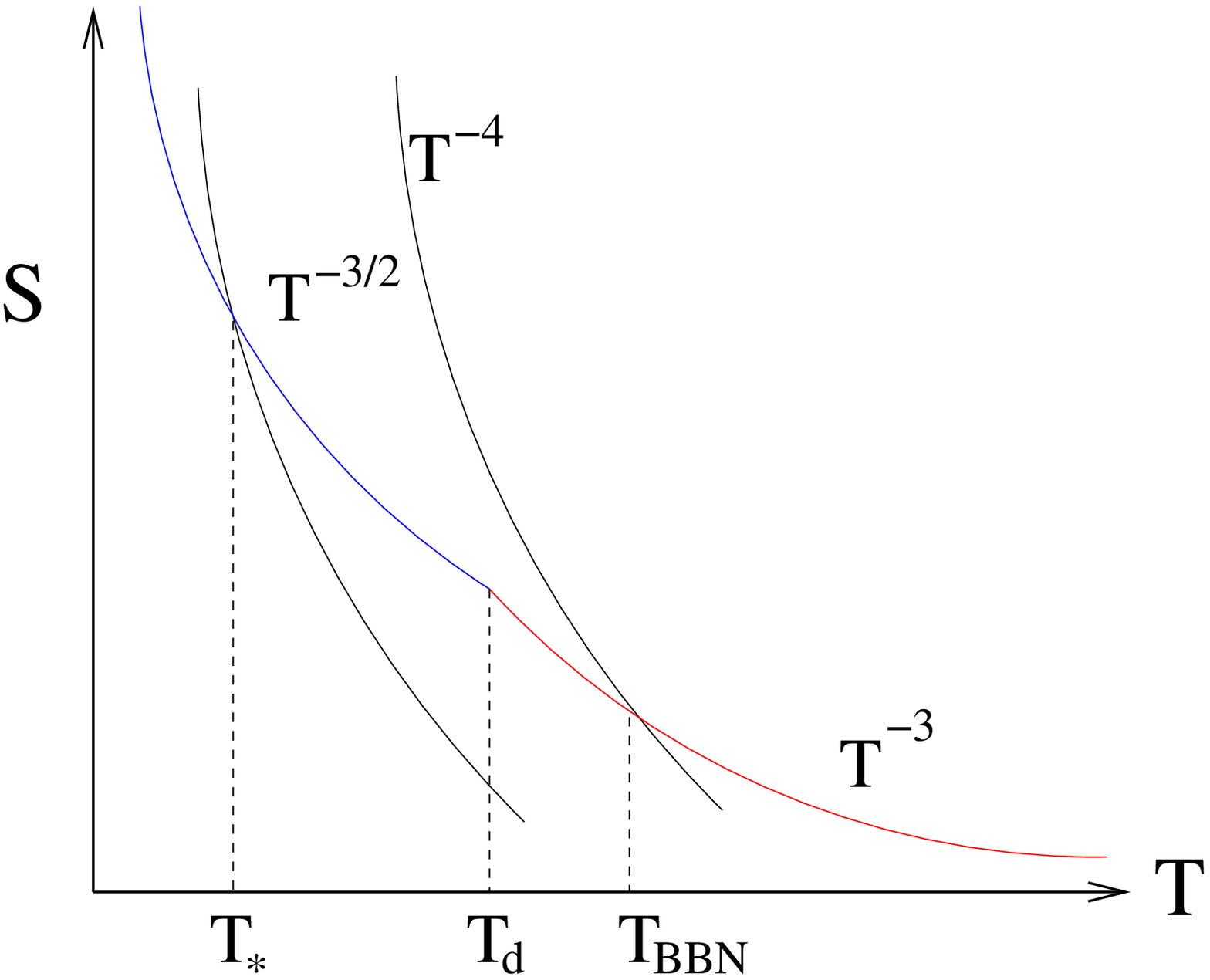}}}
\caption{Entropy in the horizon}
 \label{bbn}
\end{figure}

Consider a spacelike region V of extent $r_h$ on the 3-brane, and
compare the apparent (3+1) entropy with the holographic bound
applied to a hypersurface B which is the boundary of V. Let V have
the same shape as the brane, with thickness of order $M_*^{-1}$,
so that its surface area is of order $r_h^2$ in units of $M_*$.
(It is possible that the brane is thicker than $M_*^{-1}$, forcing
us to use a larger hypersurface with more entropy density, however
it is hard to imagine that the brane thickness is parametrically
larger than the fundamental length scale.)
Impose that this region saturate the holographic bound, so $r_h$
satisfies~\cite{foot2} (See Fig.~\ref{bbn})
\begin{equation} \label{saturate} T^3 r_h^3 \sim M_*^2 r_h^2~~,
\end{equation}
or \begin{equation}
r_h \sim T^{-1} \left( {M_* \over T} \right)^2 ~~.
\end{equation}

Now consider a cosmological horizon volume of size $d_H \sim M_P /
T^2$ (assuming radiation domination). The ratio of $r_h$ to $d_H$
is \begin{equation} \label{horizon} { d_H \over r_h} ~\sim~ {T \over M_*}{M_P
\over M_*} ~\sim~ {T \over {\rm 10^{-4} \, eV}} ~~~.\end{equation} For the
matter-dominant epoch, the horizon distance is given as $d_H\sim
\left({M_P/T_d^2}\right)\left({T_d/ T}\right)^{3/2}$, where
$T_d\simeq 10~{\rm eV}$ is the onset temperature of matter
domination. The ratio then becomes
\begin{equation}
\label{dr} {d_H\over r_h}\sim {M_P\over M_*}\left({T^3\over
M_*^2T_d}\right)^{1/2} \sim \left({T\over 10^{-2}~{\rm
eV}}\right)^{3/2}.
\end{equation}
We find  that for any temperature higher than
${\rm 10^{-2} \, eV}$ the causal horizon contains more degrees of
freedom than are allowed according to the HB applied to the
fundamental theory.

Our understanding of thermodynamics and statistical physics is
based on counting states. If the HB is correct, the early universe
in the brane worlds under consideration will likely not obey the
usual laws of thermal physics at temperatures $>{\rm 10^{-2} \,
eV}$. This makes our understanding of nucleosynthesis and the
microwave background problematic.

In order that our thermodynamic description of nucleosynthesis (at
$T \sim 10$ MeV) not be invalidated by holography, we find that
$M_* > 10^4$ TeV. (This bound is reduced slightly from
(\ref{horizon}) when prefactors in the expressions for the entropy
density and horizon size are included.)

\bigskip
\noindent  (2) {\it The holographic bound is violated by supernova
cores.}

Consider the supernova of a star of mass $M >  8 M_\odot$, which
is powered by the collapse of an iron core and leads to neutron
star or black hole formation. In this process the entropy of the
collapsed neutron star is of order one per nucleon, so the total
entropy is roughly $10^{57}$. The radius of the core is a few to
ten kilometers, so that its area ($10^{12}~{\rm cm}^2$) in $M_*$
units is only $10^{46}$, where again we take a fiducial volume of
thickness just greater than that of the brane. (As in the
cosmological case the degrees of freedom we are counting are all
confined to the brane.) Unless $M_* > 10^6$ TeV there is a
conflict between the usual thermodynamic description of supernova
collapse and the holographic entropy bound.

\bigskip
\noindent (3) {\it  Black hole entropy bound vs. covariant bound}

Susskind \cite{Susskind:1994vu} imagines a process in which a
thermodynamic system is converted into a black hole by collapsing
a spherical shell around it. Using the GSL, one obtains a bound on
the entropy of the system: $S_{matter} \leq A/4$, where $A$ is the
area of the black hole formed. This is a {\it weaker} conjecture
than the covariant bound, and has considerable theoretical support
\cite{Susskind:1994vu,Bousso:2002ju,Bekenstein:ur,Bekenstein:ax}.
In the application of the CB we are free to choose the
hypersurface B, as long as its lightsheet intersects all of the
matter whose entropy we wish to bound, whereas in the black hole
bound the area which appears is that of the black hole which is
formed. The black hole entropy bound is sensitive to the dynamics
of horizon formation.

In TeV gravity scenarios, the black hole size on the 3-brane is
controlled by the apparent Planck scale $M_P = 10^{19}$ GeV. The
extent of the horizon in the perpendicular directions off the
brane depends on the model, unless the hole is very small.

In ADD worlds, the horizon of an astrophysical black hole likely
extends to the boundary of the compact extra dimensions. As
discussed in \cite{Argyres:1998qn}, large black holes have
geometry $S^2 \times T^{D-4}$, and the horizon includes all of the
extra volume $V_w$. Due to this additional extra-dimensional
volume, the resulting entropy density is the same as in 3+1
dimensions and there is no obvious violation of any bounds.

\begin{figure}
\begin{center}
\leavevmode \epsfig{file=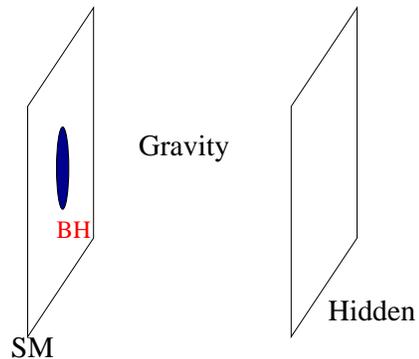,width=0.3\textwidth
} \caption{Black holes in a brane world}
\label{brane_bh}
\end{center}
\end{figure}

In RS scenarios, however, black holes are confined to the brane
and have a pancake-like geometry
\cite{Giddings:2000mu,Casadio:2002uv}. (See Fig.~\ref{brane_bh}.)
The black hole size in the
direction transverse to the brane grows only logarithmically with
the mass $M$. Thus far, no one has computed the Hawking
temperature or entropy of a pancake black hole. In fact, exact
solutions describing this objects have yet to be obtained. Let us
assume, motivated by holography, that the entropy of a pancake
black hole continues to be of order its surface area in units of
$M_*$. The surface area of a large hole is dominated by the $r^3
l^{D-5}$ component, so the black hole entropy bound arising from
the Susskind construction in RS worlds is of the form
\begin{equation}
\label{Susskind} S < (r M_*)^3~~.
\end{equation}
That is, the upper bound
on the entropy grows with the apparent 3-volume of the region. In
this case the black hole bound is clearly weaker than the
covariant bound, because the surface B used in the application of
the latter is much smaller than the area of the pancake hole.
Interestingly, (\ref{Susskind}) is the same result one would have
obtained naively from $D= 3+1$ quantum field theory in the absence
of gravity, with ultraviolet cutoff $M_*$!

\bigskip

\section{Discussion}
Our results can be interpreted in two ways, depending on how one
views holography and related entropy bounds.

It seems likely that holography is a deep result of quantum
gravity, relating geometry and information in a new way
\cite{Bousso:2002ju}. If so, it provides important constraints on
extra dimensional models. Our analysis shows that the ordinary
thermodynamic treatment of nucleosynthesis and supernovae are in
conflict with the covariant bound. In other words, brane worlds
obeying holography do not reproduce the observed big bang thermal
evolution or stellar collapse. Exactly what replaces the usual
behavior is unclear - presumably it is highly non-local - but the
number of degrees of freedom is drastically less than in the
thermodynamic description.

An alternative point of view is to regard brane worlds as a
challenge to holography. If such worlds exist they have the
potential to violate the entropic bounds by arbitrarily large
factors. However, it must be noted that the basic dynamical
assumptions underlying the scenarios (that the 3-brane and bulk
geometry arise as a ground state of quantum gravity) have never
been justified. All violations discussed here require a hierarchy
between $M_P$ and $M_*$, or equivalently that the
extra-dimensional volume factor
$V_w = \int d^{D-4}x ~\sqrt{-g_{(D-4)}}  $  exceed its
``natural'' size $\sim M_*^{-(D-4)}$.

Finally, we note that the brane, or whatever confines matter to 3
spatial dimensions, is absolutely necessary for these entropy
violations. Without the brane, matter initially in a region with
small extent in the extra $(D-4)$ dimensions will inevitably
spread out due to the uncertainty principle. For ordinary matter
in classical general relativity, in the absence of branes, Wald
and collaborators \cite{Flanagan:1999jp} have proven the covariant
entropy bound subject to some technical assumptions.

\section*{Acknowledgements}
\noindent

The work of D.K.H. is supported by KRF PBRG 2002-070-C00022.
The work of S.H. was supported in part
under DOE contract DE-FG06-85ER40224 and by the NSF through
through the USA-Korea Cooperative Science Program, 9982164.

\bigskip



\begin{thebibliography}{99}
\bibitem{foot1}If they don't, one can define the
upper bound to be the difference in the areas that bound the
null sheet~\cite{Bousso:2002ju}.
\bibitem{foot2} The area of the hypersurface of the thermal system
has a component perpendicular to the brane (blue shaded region in
Fig.~\ref{bbn}), whose area is $r_h^3$. However, there will be no entropy
flow into the lightsheet extended from this component of the hypersurface,
since all matters are confined on the brane.
\end{thebibliography}
\end{document}